# Unveiling the Nexus Between Economic Complexity and Environmental Sustainability: Evidence from BRICS-T Countries

Emre AKUSTA[1]

| Ekonomik Karmaşıklık ve Çevresel Sürdürülebilirlik Arasındaki Bağlantının Ortaya Çıkarılması: BRICS-T Ülkelerinden Kanıtlar | Unveiling the Nexus Between Economic Complexity and Environmental Sustainability: Evidence from BRICS-T Countries |
|---|---|
| **Öz** | **Abstract** |
| Bu çalışma BRICS-T ülkelerinde ekonomik karmaşıklığın çevresel performans üzerindeki etkilerini analiz etmektedir. Analizde 1999-2021 dönemi yıllık verileri, Durbin-Hausman eşbütünleşme testi ve Augmented Mean Group (AMG) tahmincisi kullanılmıştır. Panel AMG sonuçlarının sağlamlığı ise CCEMG ve CS-ARDL yöntemleri ile test edilmiştir. Sonuçlar, ekonomik karmaşıklığın çevresel performans üzerinde olumlu bir etkisi olduğunu göstermektedir. Ekonomik karmaşıklık endeksindeki %1'lik bir artış BRICS-T ülkelerinde çevresel performansı %0,020 ile %1,243 arasında artırmaktadır. Bununla birlikte, ekonomik büyüme, enerji yoğunluğu ve nüfus yoğunluğunun çevresel performans üzerinde olumsuz bir etkiye sahip olduğu bulunmuştur. Buna karşılık yenilenebilir enerji kullanımı çevresel performansa olumlu katkıda bulunmaktadır. | This study analyses the impacts of economic complexity on environmental performance in BRICS-T countries. Annual data for the period 1999-2021, Durbin-Hausman cointegration test and Augmented Mean Group (AMG) estimator are used in the analysis. The robustness of the Panel AMG results is tested with CCEMG and CS-ARDL methods. The results indicate that economic complexity has a positive impact on environmental performance. An increase of 1% in the economic complexity index increases environmental performance in BRICS-T countries between 0.020% and 1.243%. However, economic growth, energy intensity and population density were found to have a negative impact on environmental performance. Renewable energy use, in contrast, contributes positively to environmental performance. |
| **Anahtar Kelimeler:** Ekonomik Karmaşıklık, Çevresel Performans, BRICS-T, AMG, CCEMG, CS-ARDL | **Keywords:** Economic Complexity, Environmental Performance, BRICS-T, AMG, CCEMG, CS-ARDL |
| **JEL Kodları:** Q43, O44, Q56 | **JEL Codes:** Q43, O44, Q56 |

| Statement of Research and Publication Ethics | This study was prepared in accordance with the rules of scientific research and publication ethics. |
|---|---|
| Authors' Contributions to the Article | All processes of the study were conducted and completed by a single author. |
| Conflict of Interest Statement | There is no conflict of interest for the author or third parties arising from the study. |

---
[1] Assoc. Prof. Dr., Kırklareli University, Faculty of Economics and Administrative Sciences, Department of Economics, emre.akusta@klu.edu.tr

## 1. Introduction

Environmental degradation has become one of the major problems facing today's societies. Environmental problems such as climate change, air and water pollution, loss of biodiversity and depletion of natural resources have a negative impact on ecosystems. They also pose major risks to the well-being of societies, public health and economic development. Therefore, environmental sustainability is not only the basis of environmental policies, but also of economic strategies and development policies. In this regard, how countries design their economic structures and the environmental impacts of their production structures are very important (Rafique et al., 2022). Therefore, a better comprehension of the relationship between environment and economy is seen as an important step towards environmental sustainability.

There are many studies in literature analysing the relationship between environmental problems and economic development. These studies emphasise the environmental costs of economic growth. For example, the Environmental Kuznets Curve analyses the relationship between economic growth and environmental quality. It suggests that after a certain level of economic development, the negative impacts of countries on the environment will start to decrease (Grossman & Krueger, 1995). However, in practice, this relationship varies depending on many factors. Moreover, it does not give the same result for every country (Laverde-Rojas et al., 2021). Studies on the validity of the EKC hypothesis can be divided into two main groups. The first group of studies supports the validity of the EKC hypothesis. These studies show that economic growth can increase environmental sustainability after a specific stage (e.g. Gill et al., 2019; Chen et al., 2020; Yıldırım & Yıldırım, 2021). These findings confirm the theoretical framework of the inverse U-shaped relationship between economic growth and environmental quality. The second group of studies points out that the EKC hypothesis is not valid or evidence is lacking. For example, some studies found no evidence to support the EKC hypothesis (e.g. Boufateh, 2019; Li et al., 2021). Other studies have found that the EKC exhibits a different form of relationship instead of an inverse U (e.g. Arnaut & Lidman, 2021). These two groups of studies reveal that the EKC hypothesis may differ across countries, sectors and environmental factors. Therefore, more modern strategies are required to ensure environmental sustainability instead of traditional growth-oriented approaches.

The environmental impacts of production processes have become even more important in modern economies. The transformation of production processes has made the environmental impacts of modern economies more pronounced. The transition from traditional economic activities to high-tech industries leads to an increase in energy consumption and a rise in carbon emissions (Madlener & Sunak, 2011). In this context, the concept of economic complexity provides a critical theoretical framework for analyzing the relationship between economic development and environmental sustainability. Increasing economic complexity results in a shift in production structures towards sectors that require more advanced technology and high knowledge intensity. This process can increase energy demand, encouraging more intensive use of fossil fuels and causing environmental degradation. However, higher economic complexity also enables countries to allocate more resources to strategies that support environmental sustainability. For example, economies with high levels of complexity are taking important steps to reduce their environmental impact through low carbon strategies and clean technology policies (Neagu, 2019). However, with increasing complexity, energy consumption and emissions can also increase. As a result, the net impact of economic complexity on the environment is context-specific (Neagu & Teodoru, 2019).

This relationship between economic complexity and environmental sustainability may have different consequences for different countries. Economies that produce and export more complex products are competitive in sectors that require high knowledge intensity. Conversely, economies with less complex production structures usually concentrate on activities such as agricultural production or unprocessed raw materials (Swart & Brinkman, 2020). This shows that the impacts of the level of economic complexity on the environment may vary from country to country. For example, developed countries emphasise environmental protection, while developing countries adopt strategies that support growth but ignore environmental impacts (Yilanci & Pata, 2020). This dilemma shows that the impacts of economic complexity on the environment need further investigation.

With respect to environmental sustainability, the economic complexity index (ECI) has also been linked to the environmental performance of countries. Countries with low ECI levels generally produce simpler products and cause less damage to the environment. In contrast, countries with high ECI levels can cause more environmental impact by producing more sophisticated and knowledge-intensive products (Apaydın, 2020). The global economy has become complex due to the impact of modernization and technology. Therefore, the relationship between economic complexity and ecological sustainability is becoming more complicated (Lewis, 2013). Increasing economic complexity enables the development of environmentally benign production processes. This in turn leads to a reversal of negative environmental impacts (Boleti et al., 2021). For example, in the European Union, various steps have been taken towards environmental legislation since the 1970s. As a result, environmental quality has improved. The EU's environmental protection targets, low carbon strategies and renewable energy policies are examples of practices to improve environmental quality. The European Commission aims to strike a balance between economic development and environmental sustainability. For this purpose, it uses economic complexity tools in areas such as regional development, industrial innovation and competitiveness (Zengin Taşdemir & Topcu, 2024). Despite this theory, the environmental impacts of economic complexity can vary across contexts. Some countries manage to reduce their environmental impact by taking advantage of complex production structures. However, others face environmental degradation due to high energy consumption and emissions. Therefore, the analysis of the environmental impacts of economic complexity varies depending on each country's level of development and policy priorities.

As a result, the relationship between economic complexity and environmental sustainability differs between developed and developing countries. Countries with more complex production structures face the need to develop strategies to reduce their environmental impacts while achieving a competitive position in knowledge-intensive industries. Therefore, it is necessary to conduct comparative analyses across countries to understand and minimize the environmental impacts of economic complexity. In this study, the impacts of economic complexity on environmental sustainability in BRICS-T countries are analyzed. One of the main reasons for analyzing the BRICS-T countries is that they have different levels of development, development strategies and environmental programs. This diversity provides a more balanced sample, allowing for a comprehensive analysis of the relationship between economic complexity and environmental performance. For example, Brazil is a leader in environmental performance because it has a strong forest cover and strives to manage its natural resources sustainably (Alves et al., 2024). However, countries like China and India have to contend with higher energy intensity and environmental degradation (Rehman et al., 2022; Butt et al., 2023). Similarly, Russia's high level of economic complexity reflects its diversification in technology-intensive sectors, while countries like Türkiye and China are characterized by broader industrialization strategies. In this respect, Figure 1 visualizes the data for the countries.

Figure 1: EPI and ECI Scores by Country

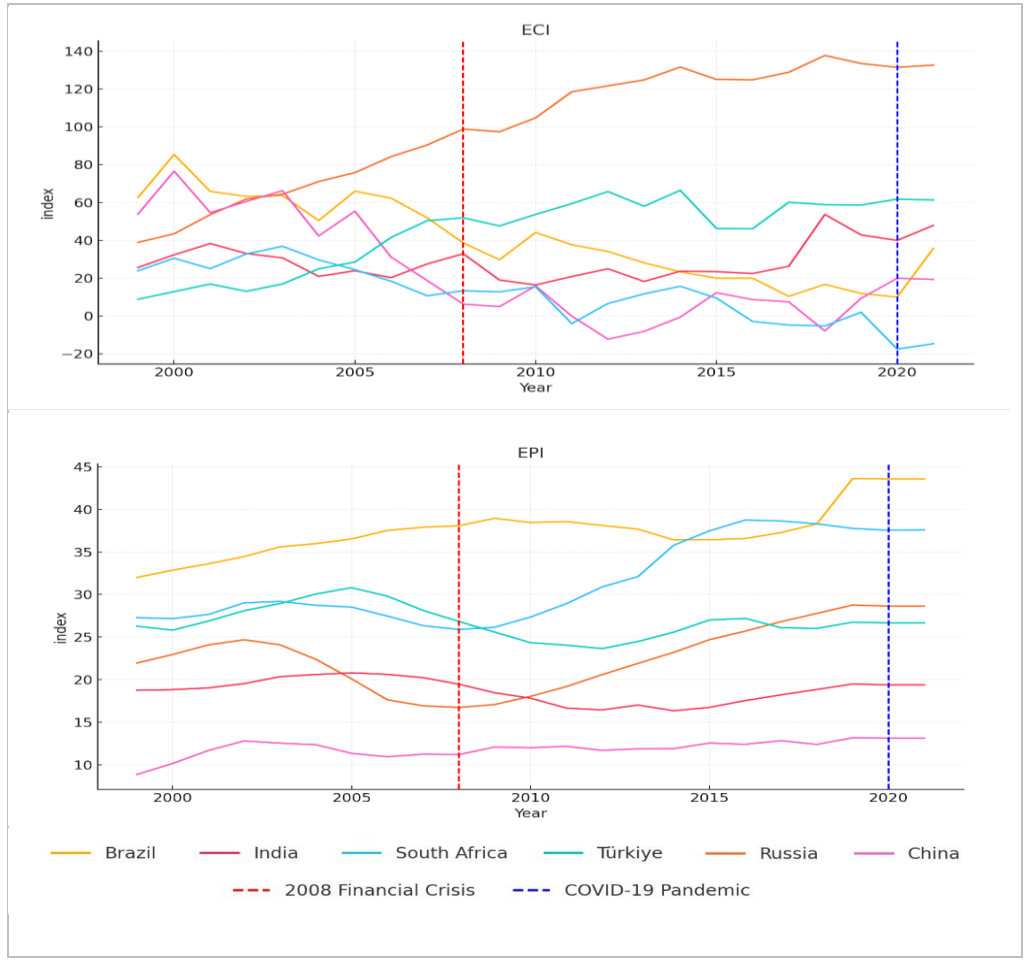

Figure 1 shows data on Environmental Performance (EPI) and Economic Complexity Index (ECI). These data clearly illustrate the differences between countries. Brazil has the highest score in terms of environmental performance with around 42 points in 2020, while countries such as India, China and Russia stand out with low scores. While China has made significant progress in improving its environmental performance between 1999 and 2020, it is still lagging behind the community. This clearly reflects the impact of countries' environmental programs and policies on environmental performance. There are also significant differences between countries in terms of the Economic Complexity Index (ECI). Russia leads in economic complexity with an index value of over 140 points in 2020, while South Africa and Türkiye have lower levels of complexity, ranging from 20-40 points. China and Brazil exhibit moderate levels of economic complexity, with an index value of around 60 points. This diversity provides an appropriate basis for further examining the differences in the economic structures of the BRICS-T countries and their impact on environmental sustainability. However, shocks such as the 2008 global financial crisis and the COVID-19 pandemic did not affect BRICS-T countries to the same degree and in the same direction. One of the main reasons for these differences is that BRICS-T countries have different levels of development, development strategies and environmental programs. For example, during the COVID-19 pandemic, Brazil exhibited relative stability in its economic and environmental performance due to its agriculture-oriented structure (Houkai & Qianwen, 2021), while China's more industrial-oriented structure was more affected by the pandemic (Lin et al., 2020). Similarly, while the 2008 crisis had a negative impact on Russia's economy based on energy exports, it had different consequences for countries with a broader industrial base such as Türkiye and South Africa. This diversity in the EPI and ECI data of the BRICS-T countries reveals that they constitute a group worth analyzing in terms of both environmental performance and economic complexity. In this respect, analyzing the role of economic complexity on environmental sustainability will contribute to the development of viable policy recommendations for achieving sustainable development goals in BRICS-T countries.

This study contributes to the literature in five main aspects. (1) Empirical studies examining the impacts of economic complexity on environmental performance in BRICS-T countries are limited. This study aims to enrich the literature in this area. (2) In many previous studies, environmental performance has been represented only by CO2 emissions. This study takes a holistic approach to environmental performance by using a more comprehensive indicator, the Environmental Performance Index. Moreover, this study goes beyond previous studies and includes the sub-dimensions of the environmental performance index in the analysis. In this scope, four different models are developed in this study. (3) Contrary to many previous studies, economic complexity is not only represented by national income. In this study, the Economic Complexity Index, which more accurately reflects the diversity and complexity of countries' production, is used. (4) Coefficient estimates are made for BRICS-T countries both overall and on a country-by-country basis. Thus, differences across countries are brought to light. (5) The study employs second generation panel data techniques that consider cross-sectional dependence. In this way, the accuracy and reliability of the analyses have been increased.

The structure of the paper is as follows: Section 2 covers the literature review, Section 3 outlines the data and methodology, Section 4 discusses the results, and Section 5 provides the conclusions.

## 2. Literature Review

The relationship between economic complexity and the environment is of increasing importance. In this respect, the existing literature addresses the impact of economic complexity on the environment from different perspectives. It also presents various findings on whether economic complexity contributes to environmental sustainability. This literature review will be presented under four sections: First, studies arguing that economic complexity has positive impacts on the environment will be reviewed. Second, studies that find that economic complexity harms the environment will be examined. Third, studies that examine the differences in the impact of economic complexity on the environment by income groups will be evaluated. Finally, studies analyzing the relationship between economic complexity and the environment within the scope of the Environmental Kuznet Curve hypothesis will be discussed. With this structure, it is aimed to provide a systematic presentation of various findings in the literature. This thematic literature structure reveals the differences and similarities between the findings. It also provides an opportunity to critically assess the impacts of economic complexity on environmental sustainability in both positive and negative ways. This approach contributes to a more structured analysis of the literature.

In the first part of the literature review, studies arguing for the positive impact of economic complexity on the environment are discussed. Among these studies, Can and Gozgor (2016) found that economic complexity reduces CO2 emissions in France. The study suggests that energy consumption has an increasing impact on environmental pollution. Despite this, it is concluded that economic complexity reduces emissions. Similarly, Azizi et al. (2019) confirmed the pollution-reducing effect of economic complexity in their study covering 99 countries. Analyzing OECD countries, Doğan et al. (2021) found that economic complexity and the use of renewable energy reduce environmental damage. These findings draw attention to the positive impact of economic structure on the environment, especially in developed economies. Boleti et al. (2021) conducted a similar study for 88 countries. The results show that economic complexity reduces environmental degradation. This confirms that economic complexity has a mitigating role on CO2 emissions. Akiş and Soyyiğit (2020) also investigated the impact of economic complexity on CO2 emissions in their study on ASEAN countries. The study confirmed the impact of economic complexity on reducing CO2 emissions, especially in countries such as Thailand, the Philippines and Indonesia.

Another study was conducted by Ikram et al. (2021) for Japan. Using the QARDL method, they proved that economic complexity has a negative impact on environmental pollution. Laverde-Rojas and Correa (2021) found that increasing economic complexity in developed economies reduces the level of environmental pollution. The results of these studies support each other. The same issue was investigated by Romero and Gramkow (2021) for 67 countries. The results revealed that increasing economic complexity reduces the intensity of greenhouse gas emissions. The study emphasized the contribution of economic complexity to environmental sustainability. Furthermore, Saqib et al. (2023) argued that economic complexity in G-10 countries is important in reducing carbon emissions by encouraging environmental innovations. These studies generally support the positive impact of economic complexity on environmental sustainability. Studies arguing that economic complexity contributes to environmental sustainability suggest that economic complexity encourages more innovative and low-carbon intensive production methods. However, these findings are generally limited to high-income economies with strong technological infrastructure. Therefore, it is critical to analyze the environmental benefits of economic complexity in a broader context to question its validity across different country groups.

In the second part of the literature review, studies argue for the negative impact of economic complexity on the environment are discussed. Among these studies, Lapatinas et al. (2019) analyzed 88 countries. As a result, they found that economic complexity negatively affects environmental quality. In the same direction, Neagu and Teodoru (2019) analyzed the European Union countries. The results of the study show that economic complexity increases environmental pollution. Especially in countries with high complexity, environmental degradation was found to be more pronounced. In the study for China, Yilanci and Pata (2020) investigated the impact of economic complexity on ecological footprint. The results supported the negative impacts of economic complexity on the environment. Kosifakis et al. (2020) examined 126 countries. The study revealed that economic complexity expands the ecological footprint and contributes to environmental degradation. Similarly, the analysis by Neagu (2020) shows that economic complexity is positively related to ecological footprint. Swart and Brinkmann (2020) find that economic complexity increases environmental damages such as forest fires in Brazil. Shahzad et al. (2021) find that economic complexity and fossil fuel consumption expand the ecological footprint in the US.

Abbasi et al. (2021) found that economic complexity increases CO2 emissions. They concluded that economic complexity harms environmental quality. Similarly, Ahmad et al. (2021) showed that economic complexity increases ecological footprint in their analysis for developing countries. Laverde-Rojas et al. (2021) reported that economic complexity is insufficient to reduce environmental pollution in Colombia. Nathaniel (2021) found that energy consumption and economic complexity increase ecological footprint and CO2 emissions in ASEAN countries. Among countries, this effect was found to be higher in Indonesia. Rafique et al. (2022) also showed that economic complexity and trade expand the ecological footprint in the long run. Analyzing MINT countries, Adebayo et al. (2022) find that economic complexity leads to a decline in environmental quality. Zengin Taşdemir and Topcu (2024) found that economic complexity contributes positively to environmental degradation. These studies generally show that economic complexity poses a risk to environmental sustainability and contributes to environmental degradation. Studies arguing for the negative impact of economic complexity on the environment show that environmental degradation is particularly pronounced in low-income and energy-intensive countries. These findings suggest that unmanaged economic complexity can increase environmental risks. However, it is also emphasized that these effects can be mitigated by appropriate policy and technological investments. Sustainability-oriented policies are therefore crucial for dealing with the environmental consequences of economic complexity.

In the third part of the literature review, studies examining the impact of economic complexity on the environment on the basis of country groups were examined. Among these studies, Doğan et al. (2019) analyzed countries in different income groups for the period 1971-2014. As a result, it is found that economic complexity increases environmental degradation in low- and middle-income countries, while it decreases it in high-income countries. In a similar study, Alvarado et al. (2021) investigated the relationship between economic complexity and ecological footprint in Latin America. They found that economic complexity expands the ecological footprint in high-income countries, while it causes less damage to the environment in low-middle-income countries. These findings suggest that the development of economic structure differentiates environmental impacts.

Neagu and Negau (2022) state that economic complexity reduces environmental pressure in developed countries. They emphasize that environmental pollution increases during the transformation process of the economic structure; however, these effects can be reversed when economic and social developments are combined with institutional supports. In this framework, economic complexity becomes a contributing factor to environmental sustainability. Adedoyin et al. (2021) examined the impact of economic complexity on CO2 emissions based on the World Bank's income classification. The results show that economic complexity increases CO2 emissions in low-income countries and decreases them in middle- and high-income countries. The validity of the EKC hypothesis is also confirmed in this study.

Finally, Ullah et al. (2024) examined the effects of economic complexity, FDI and renewable electricity load capacity factor. The analysis on BRICS-T countries finds that a positive shock in economic complexity has a positive impact on the load capacity factor in the long run, but not in the short run. Renewable electricity improves environmental quality in both periods. Among other factors, negative shocks in FDI have a positive impact on the environment in the long run. On the contrary, fossil fuel consumption and economic growth have negative impacts on the load capacity factor. These studies show that the environmental impact of economic complexity varies by country income level. Moreover, economic complexity provides more beneficial results in higher income groups. Studies show that the environmental impacts of economic complexity vary with income level. This result suggests that they are shaped not only by the level of economic complexity, but also by institutional capacities and innovation potential. Moreover, the literature shows that increasing economic complexity has more positive impacts in high-income countries and negative impacts in low-income countries. These findings emphasize that the impacts of economic complexity are not context-independent. This suggests that cross-country comparisons should consider differences in income levels.

In the fourth part of the literature review, studies that address the relationship between economic complexity and the environment within the scope of the Environmental Kuznet Curve hypothesis are reviewed. Among these studies, Neagu (2019) examined the impact of economic complexity on CO2 emissions in European Union countries within the scope of the EKC hypothesis. As a result, it was found that economic complexity exhibits an inverted-U-shaped trend on emissions. In the same study, energy intensity was found to increase CO2 emissions. Similarly, Pata (2021) tested the mitigating impact of economic complexity on environmental degradation in the US. The study revealed that the EKC hypothesis is valid and the impact of economic complexity on the environment turns positive after a certain point. Similarly, Chu (2020) analyzed 118 countries. The study found that the impact of economic complexity on the environment is in the form of an inverted-U, in line with the EKC hypothesis. That is, economic complexity is not always beneficial to the environment, but when it reaches a certain level, it reduces environmental pollution.

In a study on France, Can and Gozor (2017) examined the impact of economic complexity on CO2 emissions. The study confirmed the EKC hypothesis. For Türkiye, Özbek and Naimoğlu (2022) reported that economic complexity increases ecological footprint in the short run but decreases it in the long run. Their study shows that the EKC hypothesis is valid for Türkiye. Tekbaş and Tutumlu (2023) examined the relationship between economic complexity and CO2 emissions in Türkiye under the EKC hypothesis. However, they found that this hypothesis is not valid for Türkiye. The findings suggest that economic complexity increases CO2 emissions in the short and long run. The study also draws attention to the mitigating impact of renewable energy consumption on CO2 emissions. These studies generally argue that the impact of economic complexity on environmental pollution decreases with increasing income level. In this respect, it is argued that economic complexity contributes to environmental improvements after crossing a certain threshold. However, some examples do not confirm this relationship. Studies show that the impact of economic complexity on the environment is not linear and may exhibit an inverted-U-shaped trend. This suggests that economic complexity may initially increase environmental degradation, but may contribute to environmental improvements after a certain level of development. However, there are also examples in the literature that are not consistent with the EKC hypothesis. This emphasizes that the impacts of economic complexity may differ based on countries' income level, energy consumption profile and renewable energy utilization rates. In this respect, the contribution of economic complexity to environmental sustainability should be linked not only to economic development but also to environmental policies and the promotion of innovative technologies.

There is no consensus in the literature on the environmental impacts of economic complexity. Studies have generally found a statistically significant relationship between economic complexity and the environment. However, the direction of this relationship varies depending on the country, period and methods used. Some studies consider economic complexity as one of the main drivers of environmental degradation. It is argued that increasing economic complexity creates a challenge for countries in achieving environmental sustainability (e.g., Neagu, 2020; Shahzad et al., 2021; Rafique et al., 2022; Zengin Taşdemir & Topcu, 2024). Moreover, there are also studies suggesting that economic complexity contributes to environmental sustainability (e.g., Akiş & Soyyiğit, 2020; Doğan et al., 2021; Boleti et al., 2021; Saqib et al., 2023). Another factor that differentiates the impact of economic complexity on the environment is the income levels of countries. General findings in the literature suggest that economic complexity increases environmental degradation in low- and middle-income countries and decreases it in high-income countries (e.g., Doğan et al., 2019; Neagu, 2019; Adedoyin et al., 2021; Neagu et al., 2022), Higher environmental awareness in developed countries encourages the transition to cleaner and more advanced technologies in these countries. This is in line with the literature arguing that technological advances help protect the environment.

The literature review shows that studies that measure environmental quality by $CO_2$ emissions tend to find the impact of economic complexity on the environment to be negative. In addition, studies that equate economic complexity with economic growth show the same tendency. However, studies that look at the impact of economic complexity on the environment from a broader perspective argue that complexity contributes positively to the environment. Contrary to biases, economic complexity is not only equivalent to economic growth. The more complex and knowledge-intensive products a country produces, the higher its economic complexity. This also increases its development potential. It is an indirect measure of an economy's knowledge and capabilities. After all, the orientation of a complex economy towards knowledge-intensive technologies leads to the development of technologies that contribute to environmental sustainability in the long run.

### 3. Data and Methodology

### 3.1. Model Specification and Data

This study analyzes the impacts of economic complexity on environmental performance in BRICS-T countries. Annual data for the period 1999-2021 are used in the study. The period of the study was determined based on the availability of data. Descriptive statistics of the data set are presented in Table 1.

Table 1: Descriptive Statistics

| Variables | Symbol | Description | Mean | S. D. | Min. | Max | Source |
|---|---|---|---|---|---|---|---|
| Climate Change | PCC | index | 1.367 | 0.143 | 0.859 | 1.593 | YC |
| Environmental Health | HLT | index | 1.434 | 0.228 | 0.905 | 1.701 | YC |
| Ecosystem Vitality | ECO | index | 2.186 | 0.642 | 0.001 | 2.505 | YC |
| Environmental Performance | EPI | index | 1.829 | 0.559 | 0.001 | 2.133 | YC |
| Economic Complexity | ECI | index | 0.189 | 0.093 | 0.001 | 0.407 | HU |
| Gross Domestic Product | GDP | constant 2015 US$ | 27.849 | 1.018 | 26.083 | 30.394 | WB |
| Renewable Energy | REN | % energy consumption | 2.734 | 0.883 | 1.163 | 3.912 | WB |
| Energy Intensity | INT | MJ/$2017 PPP GDP | 1.746 | 0.445 | 0.908 | 2.49 | WB |
| Population Density | POP | per square kilometer | 4.105 | 1.251 | 2.165 | 6.16 | WB |

Note: (1) S.D., Min, and Max denote standard deviation, minimum, and maximum, respectively. (2) YC, HU, and WB indicate Yale Center for Environmental Law & Policy-Environmental performance index, Harvard University-Growth Projections and Complexity Rankings Data set, World Bank-World Development Indicators, respectively. (3) Variables are logarithmic.

The EPI is a comprehensive benchmark that assesses countries' environmental sustainability performance. The EPI assesses three main dimensions to measure countries' capacity to achieve environmental goals: (1) PCC measures countries' performance in reducing greenhouse gas emissions, lowering their carbon footprint and combating climate change. Indicators under the PCC reflect countries' GHG emission trends and efforts to reduce these emissions. (2) HLT covers environmental factors that directly affect human health. The HLT includes indicators of performance in areas such as air quality, access to clean drinking water and waste management. The impacts of air pollution, water pollution and heavy metals on human health are the main focus of this dimension. (3) ECO covers issues such as biodiversity, habitat conservation and provision of ecosystem services. ECO assesses the capacity of countries to sustainably manage their natural resources and ensure the conservation of ecosystems. Indicators such as ecosystem losses, tree cover changes, agricultural sustainability are included in this dimension. With these three dimensions, the EPI is an inclusive index that measures the environmental performance of countries. This structure enables policymakers to identify priority areas for improving environmental sustainability (Block et al., 2024).

Economic complexity refers to the amount of knowledge embedded in a country's production structure. It is measured by the diversity and complexity of products produced by countries (Hausmann et al., 2014). This index is an indicator that shows the level of development of countries and reflects their capacity to produce high value-added products (Neagu, 2020). The Economic Complexity Index (ECI) is calculated based on the diversity and complexity of products produced by a country using international trade data. The economic development processes of countries usually involve a transition from simple products to more complex products. This is only possible if developed societies produce products with more complex knowledge and export these products competitively (Hausmann et al., 2014).

Unlike traditional single-factor analyses, the ECI method brings together many components. Thus, it evaluates the development outcomes of countries from a broader perspective (Hausmann et al., 2014). The calculation of the Economic Complexity Index (ECI) follows certain steps. First, data collection and analysis of the export structure are carried out. At this stage, countries' export data are analyzed to determine which products they specialize in. This analysis reveals in which product groups each country is concentrated and the level of complexity of these products. In the second stage, the level of complexity of products is determined. Since more complex products require a broader knowledge base and various skills, the level of knowledge and skills required to produce them is analyzed. Therefore, countries that specialize in the production of complex products are considered to have more developed and complex economic structures. Finally, the index is calculated using the data obtained. This index provides a value that considers the complexity and diversity of products that a country exports. This index value is used to compare the level of economic complexity of countries (Hausmann et al., 2014).

This study analyzes the impacts of economic complexity on environmental performance in BRICS-T countries. For a more detailed analysis, the impacts on both the Environmental Performance Index (EPI) and its components, such as Climate Change (PCC), Environmental Health (HLT) and Ecosystem Vitality (ECO), are investigated separately. This decomposition allows us to more clearly observe the impact of economic complexity on different aspects of the environment.

Within the scope of the analysis, various explanatory variables that may affect environmental performance are included in the model. These variables are selected from factors commonly used in the literature due to their strong theoretical linkages with environmental challenges. The selection of control variables considers both the theoretical framework and empirical findings in the literature. First, gross domestic product (GDP) per capita is considered as an effective factor on environmental performance. GDP is important in terms of reflecting the environmental impacts of economic growth. Higher income levels may encourage increased environmental awareness and investments in clean technology. However, increased economic activity can increase energy consumption and carbon emissions. These complex dynamics are often discussed in the literature, making it important to examine the effects of GDP on environmental performance (e.g. Boleti et al., 2021; Ahmed et al., 2022; Adebayo et al., 2022). In addition, the share of renewable energy consumption in total energy consumption is also included in the model. Renewable energy is one of the important components that increase environmental sustainability. It was deemed necessary to include this variable in the model to better explain the impacts of renewable energy use on environmental performance (e.g. Boleti et al., 2021; Adebayo et al., 2022). Energy intensity is another important variable that represents the energy consumption efficiency of economic activities. Lower energy intensity is associated with improved environmental performance, while higher energy intensity is often linked to increased environmental pressures. Therefore, the inclusion of the energy intensity variable in the model allowed for the examination of the economic dimension of environmental sustainability (e.g. Ahmed et al., 2022; Ghosh et al., 2022). Finally, population density is considered as a factor that can have direct and indirect impacts on environmental pressures. High population density can increase resource use. However, environmental damage can be minimized through sustainable urbanization and infrastructure planning. For this reason, it was deemed appropriate to include the population density variable in the model, which is frequently included among the determinants of environmental performance in the literature (e.g. Boleti et al., 2021; Ahmed et al., 2022). These control variables allowed for a more comprehensive assessment of the impacts on environmental performance. The research models can be expressed as in Equations 1-4:

$$PCC_{i,t} = f(ECI_{i,t}, GDP_{i,t}, REN_{i,t}, INT_{i,t}, POP_{i,t}) \tag{1}$$

$$HLT_{i,t} = f(ECI_{i,t}, GDP_{i,t}, REN_{i,t}, INT_{i,t}, POP_{i,t}) \tag{2}$$

$$ECO_{i,t} = f(ECI_{i,t}, GDP_{i,t}, REN_{i,t}, INT_{i,t}, POP_{i,t}) \tag{3}$$

$$EPI_{i,t} = f(ECI_{i,t}, GDP_{i,t}, REN_{i,t}, INT_{i,t}, POP_{i,t}) \tag{4}$$

After logarithms are taken, the research model can be expressed as in Equation 5-8:

$$PCC_{i,t} = \beta_0 + \beta_1 ECI_{i,t} + \beta_2 GDP_{i,t} + \beta_3 REN_{i,t} + \beta_4 INT_{i,t} + \beta_5 POP_{i,t} + \varepsilon_{i,t} \tag{5}$$

$$HLT_{i,t} = \beta_0 + \beta_1 ECI_{i,t} + \beta_2 GDP_{i,t} + \beta_3 REN_{i,t} + \beta_4 INT_{i,t} + \beta_5 POP_{i,t} + \varepsilon_{i,t} \tag{6}$$

$$ECO_{i,t} = \beta_0 + \beta_1 ECI_{i,t} + \beta_2 GDP_{i,t} + \beta_3 REN_{i,t} + \beta_4 INT_{i,t} + \beta_5 POP_{i,t} + \varepsilon_{i,t} \tag{7}$$

$$EPI_{i,t} = \beta_0 + \beta_1 ECI_{i,t} + \beta_2 GDP_{i,t} + \beta_3 REN_{i,t} + \beta_4 INT_{i,t} + \beta_5 POP_{i,t} + \varepsilon_{i,t} \tag{8}$$

In Equation 5-8, $i$ = 1, …, 6 denotes each country and $t$ = 1999, …, 2021 denotes time. $\beta_0$ ve $\beta_1$ denote the intercept and error terms, while $\beta_1, \beta_2, \beta_3$ ve $\beta_4$ denote long-run elasticities.

### 3.2. Methodology

In this study, panel data analysis is employed to analyze the impacts of economic complexity on environmental performance in BRICS-T countries. For a more robust analysis, cross-sectional dependence is first tested. Then, slope homogeneity test was applied. Panel unit root tests were conducted to assess the stationarity of the variables used in the study. Cointegration analysis was applied to examine whether the variables have a long-run equilibrium relationship. Finally, coefficient estimates are made to estimate the long-run impacts of economic complexity on environmental performance.

*Cross-sectional Dependence Tests:* In panel data analysis, there may be hidden interactions between the units in the data set. These dependencies can significantly affect the results of panel data analysis. Econometric methods to be applied without horizontal cross-sectional dependence tests may produce misleading results (Chang et al., 2015). In this respect, different methods were used to test for horizontal cross-section dependence in our study: (1) Breusch-Pagan LM test (Breusch & Pagan, 1980) is used to test for the presence of cross-sectional dependence in large data sets and to assess its correlation. (2) Pesaran scaled LM test (Pesaran, 2004) is valid for smaller sample sizes and efficiently analyzes cross-sectional dependence. (3) The Bias-corrected scaled LM test (Baltagi et al., 2012), which adds bias correction, provides more robust results, especially for small sample sizes. (4) The Pesaran CD test (Pesaran, 2004) is a method for testing cross-sectional dependence in dynamic panels. It detects cross-sectional dependence more precisely regardless of the degrees of freedom.

*Slope Homogeneity Test:* In panel data analysis, it is important to determine whether all units have the same slope coefficient. Homogeneity of slope coefficients is an important factor affecting the validity of the analysis. In our study, we use the slope homogeneity test developed by Pesaran and Yamagata (2008). This test tests whether the coefficients are homogeneous and determines whether all units have the same structure. The null hypothesis assumes that all coefficients are equal, while the alternative hypothesis states that at least one coefficient is different from the others (Pesaran & Yamagata, 2008). This test is an important step towards ensuring the consistency of slope coefficients in panel data analysis.

***Panel Unit Root Tests (CIPS):*** In panel data analysis, it is important to determine the stationarity level of the series. First generation panel unit root tests may give misleading results due to the assumption of horizontal section independence (Breitung & Pesaran, 2005). Therefore, Cross-Sectional Augmented IPS (CIPS) test (Westerlund, 2008), which considers cross-sectional dependence, is preferred. This test evaluates whether the series are stationary by considering horizontal cross-sectional dependence and provides more reliable results.

***Cointegration Test:*** Panel cointegration test is used to determine whether there is a long-run relationship between variables in panel data analysis. Traditional panel cointegration tests may ignore cross-sectional dependence and heterogeneity. This may lead to limited results (Mensah et al., 2019). Therefore, the Durbin-Hausman cointegration test is used in this study. The Durbin-Hausman test (Westerlund, 2008) reliably tests the cointegration relationship in panel data sets by considering both cross-sectional dependence and heterogeneity (Amin et al., 2020). This test analyzes using two main statistics (DH Panel and DH Group). As a result, it is determined whether the variables in the panel data set have a long-run equilibrium relationship.

***Long-run Coefficient Estimation:*** This study utilizes the Augmented Mean Group (AMG) estimator to estimate the long-run coefficients. The AMG estimator is robust to problems such as cross-sectional dependence, heterogeneity, endogeneity and serial correlation. It provides consistent forecasts even under horizontal cross-section dependence. This method, developed by Eberhardt and Bond (2009), has the ability to deal with non-stationary variables. With these features, it stands out as a reliable tool in panel data analysis (Tenaw & Hawitibo, 2021). The long-run coefficients estimated separately for each unit are averaged to obtain the overall panel coefficients. This method has enabled us to obtain reliable and robust estimates by considering the heterogeneity and horizontal cross-sectional dependence in panel data.

***Robustness Methods:*** The robustness of the panel AMG results is tested with CCEMG and CS-ARDL methods. CCEMG (Common Correlated Effects Mean Group) was developed by Pesaran (2006). It is a method that considers cross-sectional dependence and heterogeneity in panel data. This method provides more robust estimation of the effects of endogenous and exogenous variables by controlling for common factor effects in the model. The long-run coefficients estimated separately for each unit are averaged to obtain the overall panel coefficients. CCEMG stands out as a reliable tool in panel data analyses as it is robust to problems such as cross-sectional dependence and heterogeneity. CS-ARDL (Cross-Sectionally Augmented ARDL) is a method used to estimate long-run relationships and control for cross-sectional dependence. This method, developed by Chudik and Pesaran (2015), extends ARDL models to account for cross-sectional dependence. CS-ARDL stands out with its ability to deal with unit root and cointegration problems in panel data sets. It provides consistent estimates by adding common factors or dynamic terms to control for horizontal cross-sectional dependence. Estimating the long-run coefficients separately for each unit and evaluating them as an overall average increases the flexibility and reliability of the method.

## 4. Results and Discussion

In the 1st stage of the analysis, cross-sectional linkages were analyzed for the variables used in the study. The results of the cross-sectional dependence test for BRICS-T countries are presented in Table 2.

Table 2: Cross-Sectional Dependence Test Results

|  | Breusch-Pagan LM | Pesaran scaled LM | Bias-corrected scaled LM | Pesaran CD |
|---|---|---|---|---|
| PCC | 49.408*** | 6.282*** | 6.146*** | 1.981*** |
| HLT | 320.599*** | 55.794*** | 55.658*** | 17.901*** |
| ECO | 81.997*** | 12.232*** | 12.096*** | 1.742*** |
| EPI | 59.089*** | 8.050*** | 7.913*** | 3.937*** |
| ECI | 158.734*** | 26.242*** | 26.106*** | 2.990*** |
| GDP | 319.430*** | 55.581*** | 55.445*** | 17.867*** |
| REN | 140.669*** | 22.944*** | 22.807*** | 7.735*** |
| INT | 196.584*** | 33.153*** | 33.016*** | 7.372*** |
| POP | 229.583*** | 39.177*** | 39.041*** | 11.664*** |

Note: The symbols ***, **, and * denote significance at the 1%, 5%, and 10% levels, respectively.

Table 2 shows that all four tests for all variables are statistically significant at the 1% significance level. This indicates that there is a strong cross-sectional dependence across countries in all variables analyzed. This dependence among BRICS-T countries indicates that developments in one country can spread rapidly to other countries. It also suggests that there may be an interdependent structure among countries. Therefore, in order to analyze the interactions among BRICS-T countries correctly, econometric methods that take horizontal cross-sectional dependence into account should be used. An approach that ignores horizontal cross-sectional dependence may negatively affect the reliability of the results of the study and may lead to bias.

Table 3: Slope Homogeneity Tests Results

| Models | Tests | LM statistics | P-value |
|---|---|---|---|
| Model 1 | $\widehat{\Delta}$ | 8.646*** | 0.000 |
|  | $\widehat{\Delta}_{adj}$ | 10.366*** | 0.000 |
| Model 2 | $\widehat{\Delta}$ | 8.691*** | 0.000 |
|  | $\widehat{\Delta}_{adj}$ | 10.420*** | 0.000 |
| Model 3 | $\widehat{\Delta}$ | 7.673*** | 0.000 |
|  | $\widehat{\Delta}_{adj}$ | 9.200*** | 0.000 |
| Model 4 | $\widehat{\Delta}$ | 9.378*** | 0.000 |
|  | $\widehat{\Delta}_{adj}$ | 11.244*** | 0.000 |

Note: The symbols ***, **, and * denote significance at the 1%, 5%, and 10% levels, respectively.

In the 2nd stage of the analysis, it is analyzed whether the long-run coefficients in BRICS-T countries are consistent with each other. For this purpose, the slope homogeneity test was conducted. The test results are presented in Table 3. These results indicate that the null hypothesis stating that the slopes are homogeneous is rejected. This shows that slope differences across countries are significant and should be considered in the analysis. Therefore, panel data methods that consider the heterogeneity in the data should be used.

Table 4: CIPS Panel Unit Root Test Results

| CIPS | Level | | First difference | |
|------|-------|---|------------------|---|
| Variable | Intercept | Intercept & Trend | Intercept | Intercept & Trend |
| PCC | -0.536 | -1.612 | -3.037*** | -4.083*** |
| HLT | -1.547 | -1.288 | -2.900*** | -3.178*** |
| ECO | -1.214 | -1.679 | -3.288*** | -3.380*** |
| EPI | -0.885 | -1.273 | -3.935*** | -4.004*** |
| ECI | -1.218 | -1.793 | -4.647*** | -4.981*** |
| GDP | -1.893 | -2.330 | -3.292*** | -3.320*** |
| REN | -2.243 | -2.031 | -3.890*** | -4.101*** |
| INT | -2.115 | -1.828 | -4.060*** | -4.330*** |
| POP | -2.747 | -2.748* | -4.285*** | -4.416*** |

Note: The symbols ***, **, and * denote significance at the 1%, 5%, and 10% levels, respectively.

In the 3rd stage of the analysis, the stationarity of the variables is analyzed with the CIPS panel unit root test. This test provides more reliable results since it takes horizontal cross-section dependence into account. The findings presented in Table 4 indicate that none of the variables are stationary at level. However, the variables became stationary after taking their first differences. This result means that all series are I(I). Therefore, it is possible to construct econometric models based on these stationary series.

Table 5: Durbin-Hausman Panel Cointegration

| Statistic | Model 1 | Model 2 | Model 3 | Model 4 |
|-----------|---------|---------|---------|---------|
| DH$_G$ Statistics | 9.943*** | 2.845** | 4.210* | 8.057*** |
| DH$_P$ Statistics | 6.392*** | 2.729** | 3.546* | 7.138*** |
| Decision | Cointegration | Cointegration | Cointegration | Cointegration |

Note: The symbols ***, **, and * denote significance at the 1%, 5%, and 10% levels, respectively.

In the 4th stage of the study, the long-run relationship between the variables was analyzed. For this purpose, Durbin-Hausman cointegration test was applied. Table 5 shows the results of this test. Both Durbin-Hausman group statistic (DHG) and Durbin-Hausman panel statistic (DHP) are considered in the test results. The results show that cointegration is found in all models at the 10% significance level. This shows that the variables analyzed in BRICS-T countries are in a long-run equilibrium relationship. The test results reveal that long-run changes in one variable can also affect other variables.

Table 6: Panel AMG Results for All Countries

| Variables | Model 1 | Model 2 | Model 3 | Model 4 |
|-----------|---------|---------|---------|---------|
| ECI | 0.392** | 0.020** | 1.028* | 1.243*** |
| GDP | -1.484*** | -0.423** | -1.205** | -0.960** |
| REN | 0.948*** | 0.565* | 0.175 | 0.507** |
| INT | -0.395** | -0.842** | -0.420** | -0.659** |
| POP | -0.340* | -1.498*** | -0.248*** | -1.035* |

Note: The symbols ***, **, and * denote significance at the 1%, 5%, and 10% levels, respectively.

In the 5th stage of the study, long-run coefficients are estimated. For this purpose, the panel AMG method was used and the results are presented in Table 6. The results indicate that economic complexity has an overall positive impact on environmental performance and its dimensions. The sign of ECI is positive and statistically significant. An increase of 1% in ECI leads to an improvement in environmental performance and its dimensions between 0.020% and 1.243% (Models 1-4). This finding suggests that economic complexity contributes to environmental performance. In particular, the 1.243% impact in Model 4 demonstrates that economic complexity may have a stronger positive impact on environmental performance.

The sign of GDP is negative and statistically significant. An increase of 1% in GDP causes a decrease between 0.423% and 1.484% on EPI. This suggests that economic growth has a negative impact on environmental performance. In particular, the negative impact of 1.484% observed in Model 1 suggests that economic growth may increase the pressure on PCC. Similarly, INT and POP also have a negative impact on environmental performance. An increase of 1% in INT reduces EPI by between 0.395% and 0.842%. Similarly, an increase of 1% in POP decreases EPI by 0.248% to 1.498%. These findings suggest that energy intensity and population density have a negative impact on environmental sustainability. Energy efficiency and population control can play an important role in improving environmental performance.

An increase of 1% in REN increases the EPI by 0.175% to 0.948%. This result supports that the use of renewable energy contributes positively to environmental performance. In particular, the positive impact of 0.948% in Model 1 suggests that renewable energy sources can play an important role in terms of PCC in BRICS-T countries. In conclusion, while economic complexity contributes positively to environmental performance in BRICS-T countries, economic growth poses risks to the environment. These findings emphasize that factors such as economic complexity, energy intensity and population density should be considered when developing environmental sustainability policies in BRICS-T countries. However, in order to reach more accurate results, it is important to consider the unique characteristics of each country. To this end, country-level estimates are discussed in the next stage of the analysis.

Table 7: Diagnostic Test

| Multicollinearity | | | | Heteroscedasticity (Breusch and Pagan, 1979) | Serial Correlation (Jochmans and Verardi, 2019) | Normality (Jarque and Bera, 1987) | Omitted Variable Bias (Ramsey, 1969) |
|---|---|---|---|---|---|---|---|
| Variables | VIF | 1/VIF | | | | | |
| ECI | 2.577 | 0.388 | Model 1 | 0.45 (0.620) | 41.200 (1.000) | 3.100 (0.310) | 0.65 (0.480) |
| GDP | 4.308 | 0.232 | Model 2 | 0.38 (0.710) | 43.500 (1.000) | 2.870 (0.290) | 0.68 (0.530) |
| REN | 1.755 | 0.570 | Model 3 | 0.42 (0.550) | 42.100 (1.000) | 3.050 (0.275) | 0.63 (0.610) |
| INT | 2.340 | 0.427 | Model 4 | 0.50 (0.370) | 39.300 (1.000) | 2.940 (0.360) | 0.70 (0.490) |
| POP | 4.717 | 0.212 | | | | | |

Note: p-values are given in parentheses.

In the 6th stage of the study, diagnostic tests were conducted to test the stability of the model's output. The findings are given in Table 8. The results of the multicollinearity analysis revealed that there is not a high correlation between the independent variables. The fact that VIF values are below 5 for all independent variables proves that the model estimates are reliable in this respect. In addition, the results of the variance test clearly show that the error terms have a constant variance and there is no heteroscedasticity problem. The serial correlation test results show that the error terms are independent of each other, while the normality test results indicate that the error terms have a normal distribution. These findings are important as they show that the model works in line with the basic assumptions. Moreover, the omitted variable bias test results also support that the model maintains its structural integrity.

Table 8: Robustness Results of CCEMG and CS-ARDL

| CCEMG Variables | Model 1 | Model 2 | Model 3 | Model 4 |
|---|---|---|---|---|
| ECI | 0.318** | 0.014* | 0.948** | 1.102* |
| GDP | -1.189*** | -0.354* | -1.023** | -0.867** |
| REN | 0.812*** | 0.447* | 0.137 | 0.413* |
| INT | -0.312** | -0.745* | -0.346* | -0.587* |
| POP | -0.287* | -1.189*** | -0.201** | -0.876* |
| CS-ARDL Variables | Model 1 | Model 2 | Model 3 | Model 4 |
| ECI | 0.458*** | 0.026** | 1.153*** | 1.308*** |
| GDP | -1.617*** | -0.442** | -1.297*** | -1.043** |
| REN | 1.042*** | 0.485 | 0.204 | 0.538*** |
| INT | -0.441*** | -0.894** | -0.493** | -0.692** |
| POP | -0.376** | -1.547 | -0.273*** | -1.139*** |

Note: The symbols ***, **, and * denote significance at the 1%, 5%, and 10% levels, respectively.

In the 7th stage of the study, the robustness of the panel AMG results is tested with CCEMG and CS-ARDL methods. Table 8 presents the long-run results of CCEMG and CS-ARDL[2] methods for robustness check. The results of panel AMG, CCEMG and CS-ARDL methods show generally similar trends; however, some differences are also noteworthy. In terms of similarities, in all three methods, the ECI variable has a positive and significant impact on environmental performance indicators in all models. This consistently confirms that economic complexity contributes to environmental sustainability. Similarly, the GDP variable has a negative and significant effect in all methods. These results suggest that economic growth may negatively affect environmental sustainability. REN and INT variables also show generally consistent effects across methods. The POP variable has a significant negative effect in most models and exhibits a similar trend across methods. The differences are notable in that the results of the CCEMG method show lower effects compared to AMG. For instance, the coefficients of ECI are lower in CCEMG than in AMG in all models, but they are still positive and significant. Similarly, the negative impact of GDP on environmental performance is weaker in CCEMG. This suggests that CCEMG tends to provide more conservative estimates.

Moreover, the results of the CS-ARDL method produce higher coefficients compared to the AMG results. In particular, the ECI and GDP impacts in Model 3 and Model 4 are significantly stronger in CS-ARDL compared to AMG. However, REN and POP variables are not significant in Model 2 results of CS-ARDL. This indicates that the impact of renewable energy consumption and population density on environmental performance in Model 2 may differ across methods. It can be argued that CS-ARDL provides more flexible results in this context and emphasizes the importance of model-specific variables. In conclusion, AMG, CCEMG and CS-ARDL methods show a general consistency and confirm the positive relationship between economic complexity and environmental performance. However, due to the nature of the methods, there are some differences in the estimated coefficients and these differences point to the strengths and weaknesses of the methods. Robustness analysis confirms the reliability of the long-run coefficients.

---

[2] The CS-ARDL method provides both short-run and long-run coefficient estimates. However, this study only considers long-run coefficients to facilitate comparison with panel AMG and CCEMG methods.

Table 9: Panel AMG Results for Model 1

| Variables | Brazil | Russia | India | China | S. Africa | Türkiye |
|-----------|--------|--------|-------|-------|-----------|---------|
| ECI | -0.117 | 0.421** | 0.360*** | 1.988*** | 0.052** | 0.105*** |
| GDP | -0.776*** | -0.838** | -0.994** | -0.137* | -0.597* | -1.515** |
| REN | 0.357*** | -0.004 | 0.127 | 0.850** | -0.318* | 0.568*** |
| INT | -0.264*** | -0.473** | -0.770*** | -0.991* | -0.674** | -0.944* |
| POP | -0.195* | -0.059*** | -0.923* | -1.673* | 0.859** | -0.165** |

Note: The symbols ***, **, and * denote significance at the 1%, 5%, and 10% levels, respectively.

In the 8th stage of the study, country-specific coefficients are estimated. Table 9 shows the panel AMG estimation results of Model 1. PCC measures performance on climate change. Therefore, it represents greenhouse gas emissions, carbon emissions and climate change performance of countries. In Brazil, the impact of economic complexity on climate change performance is not statistically significant. However, in Russia, India, South Africa and Türkiye, the impact of economic complexity on climate change performance is positive at significance levels between 1% and 5%. In these countries, an increase of 1% in ECI increases PCC by between 0.052% and 1.988%. This suggests that complex production structures in these countries can improve climate change performance. The high coefficient in China (1.988%) indicates that its technological infrastructure and advanced production skills contribute to environmental sustainability.

In all BRICS-T countries, the impact of GDP on climate change performance is statistically significant and negative. An increase of 1% in GDP reduces PCC by 0.137% to 1.515% in these countries. This impact is more pronounced in high-population countries such as Türkiye and India. In these countries, economic growth has a more negative impact on environmental performance. This suggests that economic activity can increase environmental costs and lead to higher emission levels. Moreover, renewable energy use has a positive and statistically significant impact on climate change performance in Brazil, China and Türkiye, but not in Russia, India and South Africa. In Brazil and Türkiye, an increase of 1% in REN increases PCC by 0.357% and 0.568%, respectively. This suggests that the use of renewable energy can play an important role in improving environmental sustainability performance.

In all countries, the impact of energy intensity on climate change performance is negative and significant. In these countries, an increase of 1% in INT reduces PCC by between 0.264% and 0.991%. This finding suggests that energy-intensive sectors have a negative impact on environmental performance and that energy efficiency policies are needed. However, the impact of population density on climate change performance is negative and significant in Brazil, India, China and Türkiye, while it is positive in Russia. In India and China, an increase of 1% in POP reduces PCC by between 0.923% and 1.673%. In this case, environmental pressure is higher in densely populated countries. Moreover, the impact of population growth on environmental performance is negative.

Table 10: Panel AMG Results for Model 2

| Variables | Brazil | Russia | India | China | S. Africa | Türkiye |
|-----------|--------|--------|-------|-------|-----------|---------|
| ECI | 0.192** | 0.819** | -0.114 | 0.814** | 0.575** | 0.476* |
| GDP | -0.501*** | -1.032* | -0.184** | 0.450*** | -0.071*** | -1.271*** |
| REN | -0.180 | 0.767* | 0.495*** | -0.722 | 0.840 | 0.808*** |
| INT | 0.882 | -0.276** | -0.388** | 0.058 | -0.548** | -0.781** |
| POP | -1.359*** | 0.977 | -0.579*** | -1.242*** | -0.379*** | -1.593*** |

Note: The symbols ***, **, and * denote significance at the 1%, 5%, and 10% levels, respectively.

Table 10 shows the panel AMG estimation results of Model 2. HLT includes environmental health indicators such as air quality, sanitation, drinking water safety, heavy metals and waste management. The impact of economic complexity on environmental health performance is significant in all countries except India. It is statistically significant at least at the 10% level in Brazil, Russia, China, South Africa and Türkiye. In these countries, an increase of 1% in ECI leads to an improvement in HLT between 0.192% and 0.819%. The particularly high coefficient in China (0.814%) suggests that the country's advanced manufacturing infrastructure and complex economic structure may have a positive impact on environmental health.

In all BRICS-T countries, the impact of GDP on environmental health performance is negative and significant. An increase of 1% in GDP causes a decrease between 0.071% and 1.271% on HLT in these countries. The fact that this impact is 1.271% in Türkiye points to the major negative impacts of growth processes on environmental health. However, in China, the impact of GDP on environmental health is positive and significant (0.450%). This shows that China has been able to achieve improvements in environmental health during the economic growth process. Moreover, renewable energy has a positive and significant impact on environmental health performance in Russia, India and Türkiye. In these countries, an increase of 1% in REN leads to an improvement in HLT between 0.495% and 0.808%. In Brazil and China, renewable energy has no significant impact on environmental health.

The impact of energy intensity on environmental health performance is negative and significant in Russia, India, South Africa and Türkiye. An increase of 1% in INT decreases HLT between 0.276% and 0.781%. This result shows that environmental health is negatively affected in countries with low energy efficiency. In Brazil and China, the impact of energy intensity on environmental health performance is not significant. Moreover, the impact of population density on environmental health performance is negative and significant in all BRICS-T countries. In these countries, an increase of 1% in POP reduces HLT by between 0.379% and 1.593%. The -1.593% impact in Türkiye indicates that the pressure of population growth on environmental health is high. In Russia, no significant impact of population density on environmental health was found.

Table 11: Panel AMG Results for Model 3

| Variables | Brazil | Russia | India | China | S. Africa | Türkiye |
|-----------|--------|--------|-------|-------|-----------|---------|
| ECI | -0.897 | 0.279** | 0.315** | 1.468*** | 0.187** | 0.383* |
| GDP | -1.423*** | -0.163** | -0.119* | -0.757* | -0.337* | 0.770 |
| REN | 0.060 | 0.087* | -0.768 | 0.060* | -0.324 | -0.062 |
| INT | -0.628*** | 0.748 | -0.575*** | -0.082*** | 0.412 | -0.395*** |
| POP | -0.170** | -0.455** | -0.296** | -1.197** | 0.758 | -0.130** |

Note: The symbols ***, **, and * denote significance at the 1%, 5%, and 10% levels, respectively.

Table 11 shows the panel AMG estimation results of Model 3. ECO includes environmental aspects such as biodiversity, habitat conservation, forest loss, fisheries and agricultural sustainability. In Brazil, the impact of economic complexity on ecosystem vitality performance is not significant. In Russia, India, China, South Africa and Türkiye, ECI has a positive impact at least at the 10% significance level. In these countries, an increase of 1% in ECI increases ECO by between 0.187% and 1.468%. The high coefficient in China (1.468%) suggests that its advanced economic structure can play a positive role in environmental sustainability.

The impact of GDP on ecosystem vitality performance is negative and statistically significant in Brazil, Russia, India, China and South Africa. An increase of 1% in GDP reduces ECO by between 0.119% and 1.423% in these countries. In particular, the -1.423% impact in Brazil suggests that economic growth puts negative pressure on natural resources and biodiversity. In Türkiye, GDP has no significant impact on ECO. Renewable energy use has a positive and statistically significant impact on ecosystem vitality performance in Russia and China. In Russia and China, an increase of 1% in REN increases ECO by 0.087% and 0.060%, respectively. However, the impact of renewable energy on ECO is not significant in Brazil, India, South Africa and Türkiye.

The impact of energy intensity on ecosystem vitality performance is negative and significant in Brazil, India, China and Türkiye. In these countries, an increase of 1% in INT reduces ECO by between 0.082% and 0.628%. The negative impacts in India and Türkiye show the negative impact of energy-intensive sectors on the ecosystem. Similarly, the impact of population density on ecosystem vitality performance is negative and significant in Brazil, Russia, India, China and Türkiye. However, no significant impact is found in South Africa. In these countries, an increase of 1% in population density reduces ECO by between 0.130% and 1.197%. The high impact of -1.197% in China indicates the pressure of dense population on ecosystems.

Table 12: Panel AMG Results for Model 4

| Variables | Brazil | Russia | India | China | S. Africa | Türkiye |
|-----------|--------|--------|-------|-------|-----------|---------|
| ECI | 0.802** | 0.826** | 0.186** | 0.074*** | 0.913* | 0.685** |
| GDP | -0.877*** | -0.080*** | -0.525*** | -0.999*** | -0.443*** | -1.494*** |
| REN | 0.444* | -0.851 | 0.651** | 0.083** | -0.009 | 0.173** |
| INT | -0.207** | 0.958 | -0.491** | -0.979** | -0.832* | -0.401* |
| POP | -0.697* | -0.878* | 0.730 | -0.884* | 0.329 | -0.579* |

Note: The symbols ***, **, and * denote significance at the 1%, 5%, and 10% levels, respectively.

Table 12 shows the panel AMG estimation results of Model 4. EPI includes environmental performance indicators such as climate change, environmental health and ecosystem vitality. The results show that economic complexity has a positive and statistically significant impact on environmental performance in all BRICS-T countries. In these countries, an increase of 1% in ECI increases EPI by between 0.074% and 0.913%. This suggests that economic complexity and technological capacity can support environmental sustainability. Developing sectors with complex production structures is important for environmental performance, especially in natural resource-based economies such as South Africa and Russia. In countries with high economic complexity, environmentally friendly production technologies and green energy are common. This also contributes to the environment.

On the contrary, the impact of GDP on environmental performance is negative and significant in all countries. An increase of 1% in GDP reduces the EPI by between 0.080% and 1.494%. The particularly high negative impact in Türkiye (-1.494%) suggests that environmental costs increase in fast-growing economies and resource consumption damages the ecosystem. Economic growth processes such as industrialization and urban sprawl lead to overuse of resources and increased emissions. Politically, the lack of environmental regulations and limited use of environmentally friendly technologies can increase the negative impacts of economic growth on the environment. Environmental policies need to be strengthened for sustainable growth. In contrast, the use of renewable energy has been found to have a positive and significant impact on environmental performance in Brazil, India, China and Türkiye. An increase of 1% in REN increases EPI by 0.083% to 0.651% in these countries. This result demonstrates the importance of investments in renewable energy sources for environmental sustainability. The high impact in India (0.651%) shows the positive impact of the transition to renewable energy on the environment. Socially, the transition to renewable energy supports public health by improving energy security and air quality. Political support and incentives for renewable energy can also accelerate this process.

Energy intensity has a negative impact on environmental performance in all BRICS-T countries. An increase of 1% in INT reduces EPI by between 0.207% and 0.979%. The high negative impact in China (-0.979%) reveals the detrimental impact of intensive energy consumption on environmental performance. Fossil fuels used in industries with low energy efficiency negatively affect environmental performance. Economically, modernizing energy-intensive sectors and introducing energy efficiency measures will support environmental sustainability. Politically, promoting energy efficiency incentives and low-carbon energy sources can have a positive impact on the environment. Similarly, the impact of population density on environmental performance is negative and statistically significant in Brazil, Russia, China and Türkiye. An increase in population density by 1% reduces EPI by between 0.579% and 0.884%. This indicates that infrastructure and resource management in metropolitan cities may be inadequate and environmental problems may increase. Especially in countries with rapidly growing populations such as Türkiye, sustainable use of environmental resources is becoming more difficult. From an environmental perspective, high population density leads to problems such as overuse of natural resources, water scarcity and air pollution. From a social perspective, dense populations have a negative impact on environmental health and quality of life. It is therefore important to stabilize the rate of population growth and develop environmentally friendly urban policies.

These results suggest that economic complexity has a positive impact on environmental performance in BRICS-T countries. The findings are consistent with studies such as Akiş and Soyyiğit (2020), Boleti et al. (2021), Adedoyin et al. (2021) and Ullah et al. (2024). However, it was also found that economic growth, energy intensity and population density negatively impact environmental performance. These results are also consistent with the literature (e.g., Boleti et al., 2021; Ahmed et al., 2022; Ghosh et al., 2022; Ullah et al., 2024). However, it is noteworthy that this impact varies across the components of the environmental performance index. For example, the impact of economic complexity on climate change is very high in China due to its advanced technology and production infrastructure. However, this impact is more limited in countries such as India and Türkiye. This difference can be explained by countries' level of technological sophistication, renewable energy utilization rates and diversity in environmental policy practices. The negative impact of GDP on environmental performance also varies across countries. For example, the higher impact in rapidly growing economies such as Türkiye can be attributed to the increased environmental costs of economic activities and resource consumption. In contrast, the negative impact of economic growth on environmental performance is more limited in China. This may be due to improved environmental regulations and energy efficiency policies. It is also noteworthy that energy intensity and population density have a negative impact on environmental performance. However, the variation in this impact across countries can be attributed to differences in energy consumption patterns and infrastructure capacity. For example, the high energy intensity in China can be explained by the prevalence of sectors with low energy efficiency, while the low impact in Brazil can be attributed to the fact that it is an economy with less energy-intensive sectors. The positive contribution of renewable energy use to environmental performance also varies across countries. Strong positive impacts in India and Türkiye can be attributed to the increase in renewable energy investments in recent years. However, the limited effects in Russia and South Africa can be attributed to their fossil fuel-based energy policies. The interaction between socioeconomic factors and environmental sustainability is important. This importance highlights the importance of environmentally friendly energy use, energy efficiency and sustainable urbanization policies in BRICS-T countries. In particular, the promotion of renewable energy sources and the modernization of energy-intensive sectors are critical steps to achieve sustainable development goals. These findings suggest that BRICS-T countries need to balance economic growth and sustainability while achieving environmental goals. Moreover, these differences emphasize that environmental sustainability policies should be designed and implemented considering the specific conditions of each country.

## 5. Conclusion

This study analyzes the impacts of economic complexity on environmental performance in BRICS-T countries. For this purpose, annual data for the period 1999-2021 and second generation panel data methods are applied. The Durbin-Hausman cointegration test is used to test the existence of long-run relationships among variables. Long-run coefficients are estimated with the Augmented Mean Group (AMG) estimator. These methods allowed us to examine in detail the differential impacts of economic complexity, growth, energy intensity, renewable energy use and population density on environmental performance.

The findings of the study show that economic complexity has a positive impact on environmental performance and its three pillars. In all models, economic complexity promotes environmental sustainability. Increased economic complexity offers the potential to reduce greenhouse gas emissions and environmental health risks by promoting the use of environmentally friendly technologies. This emphasizes the need for environmentally friendly technologies in the economic development processes of BRICS-T countries. In contrast, the impact of GDP on environmental performance is found to be negative. In all models, economic growth leads to a decline in environmental performance, thereby revealing the environmental costs of economic activities. Rapid economic growth increases resource consumption and puts pressure on natural areas. These results suggest that the growth process in BRICS-T countries should be carefully managed against environmental impacts. The use of renewable energy is found to have positive impacts on environmental performance and environmental health in particular. This finding suggests that renewable energy investments contribute to environmental sustainability. It also suggests that renewable energy sources can play a critical role for environmental sustainability in BRICS-T countries. In particular, the positive effects of renewable energy use on environmental health and ecosystem vitality suggest that further investments in this area should be encouraged. On the contrary, energy intensity and population density variables have a negative impact on environmental performance. High energy consumption and rapid population growth put significant pressure on environmental resources. This situation makes it difficult to achieve sustainable development goals in BRICS-T countries. In this regard, energy efficiency and sustainable urbanization are of great importance for improving environmental performance.

In conclusion, this study shows that economic complexity contributes positively to environmental performance in BRICS-T countries. However, economic growth, energy intensity and population density are found to be risks for the environment. Promoting renewable energy use and energy efficiency policies are important for environmental performance in these countries. The findings emphasize that BRICS-T countries need to balance economic growth with environmental sustainability. To achieve this balance, each country needs to develop policies that consider its unique characteristics. Therefore, a series of policy recommendations have been prepared: (1) Economic complexity is an element that should be supported due to its positive impacts on environmental performance. In this perspective, BRICS-T countries should invest in high value-added production and knowledge-intensive sectors. Promoting environmentally friendly technologies in production and encouraging research and development activities will positively affect environmental performance. Creating a knowledge-based economy, especially in advanced technology and innovative production areas, will support environmental sustainability. (2) The study shows that economic growth has a negative impact on environmental performance. Therefore, a sustainable growth strategy should be adopted. Instead of supporting growth in sectors with high environmental costs, it is important to promote environmentally friendly and sustainable sectors. At the same time, integrating environmental regulations into economic activities will contribute to environmental sustainability in the long run. (3) Renewable energy is an important factor contributing to environmental performance. In this direction, BRICS-T countries should increase investments in renewable energy and reduce dependence on fossil fuels. In particular, the use of renewable resources such as solar, wind and biomass energy should be encouraged. At the same time, modernizing energy infrastructure and increasing renewable energy generation capacity will support energy security and environmental sustainability together. (4) Given that energy intensity negatively impacts environmental performance, improving energy efficiency is of crucial importance. Energy efficiency programs and the use of low carbon technologies should be encouraged. Investments should be made in energy efficiency, especially in the industrial and transportation sectors. In addition, technologies to increase energy savings should be supported. This transformation can be accelerated through tax incentives and subsidies in energy-intensive sectors. (5) Population density is another important factor affecting environmental performance. In countries with rapid population growth, sustainable urbanism and infrastructure investments should be prioritized. In urban areas, measures such as protection of green spaces and water resources and waste management will protect environmental health. At the same time, sustainable urbanization and environmentally friendly infrastructure investments can reduce the negative environmental impacts of population density.

This study provides important findings on the impacts of economic complexity on environmental performance in BRICS-T countries. However, the study also has some limitations. Future research can address these limitations. First, the study only examined the BRICS-T countries. Adding other country groups would allow us to assess the impacts of economic complexity on environmental performance in different regions from a broader perspective. Second, this study is limited to the period 1999-2021. Analyses using larger data sets may provide more comprehensive information on how the relationship between economic complexity and environmental performance has changed over time. Finally, this study could not analyze structural breaks in detail. Adding different analysis methods can strengthen the consistency of the results. The use of various statistical techniques can provide a deeper insight into the relationship between economic complexity and environmental performance.

**Extended Summary**
**Unveiling the Nexus Between Economic Complexity and Environmental Sustainability: Evidence from BRICS-T Countries**

This study analyzes the impacts of economic complexity on environmental performance in BRICS-T countries. For this purpose, annual data for the period 1999-2021 and second generation panel data methods are applied. The Durbin-Hausman cointegration test is used to test the existence of long-run relationships among variables. Long-run coefficients are estimated with the Augmented Mean Group (AMG) estimator. These methods allowed us to examine in detail the differential impacts of economic complexity, growth, energy intensity, renewable energy use and population density on environmental performance.

The EPI is a comprehensive benchmark that assesses countries' environmental sustainability performance. The EPI assesses three main dimensions to measure countries' capacity to achieve environmental goals: (1) PCC measures countries' performance in reducing greenhouse gas emissions, lowering their carbon footprint and combating climate change. Indicators under the PCC reflect countries' GHG emission trends and efforts to reduce these emissions. (2) HLT covers environmental factors that directly affect human health. The HLT includes indicators of performance in areas such as air quality, access to clean drinking water and waste management. The impacts of air pollution, water pollution and heavy metals on human health are the main focus of this dimension. (3) ECO covers issues such as biodiversity, habitat conservation and provision of ecosystem services. ECO assesses the capacity of countries to sustainably manage their natural resources and ensure the conservation of ecosystems. Indicators such as ecosystem losses, tree cover changes, agricultural sustainability are included in this dimension. With these three dimensions, the EPI is an inclusive index that measures the environmental performance of countries.

Economic complexity refers to the diversity of knowledge, skills and technology components in the goods and services produced by a country or region. The ECI assesses the complexity of an economy based on the diversity and complexity of the products produced by that economy. Thus, it aims to explain the differences in prosperity between countries by considering the diversity of their production and export structures and the complexity of these structures. The more complex and knowledge-intensive products a country produces, the higher its development potential is considered to be. This approach is an indirect measure of an economy's knowledge and capabilities. Unlike traditional single-factor analyses, this method brings together many components.

The calculation of the Economic Complexity Index (ECI) follows certain steps. First, data collection and analysis of the export structure are carried out. At this stage, countries' export data are analyzed to determine which products they specialize in. This analysis reveals in which product groups each country is concentrated and the level of complexity of these products. In the second stage, the level of complexity of products is determined. Since more complex products require a broader knowledge base and various skills, the level of knowledge and skills required to produce them is analyzed. Therefore, countries that specialize in the production of complex products are considered to have more developed and complex economic structures. Finally, the index is calculated using the data obtained. This index provides a value that considers the complexity and diversity of products that a country exports. This index value is used to compare the level of economic complexity of countries.

The findings of the study show that economic complexity has a positive impact on environmental performance and its three pillars. In all models, economic complexity promotes environmental sustainability. Increased economic complexity offers the potential to reduce greenhouse gas emissions and environmental health risks by promoting the use of environmentally friendly technologies. This emphasizes the need for environmentally friendly technologies in the economic development processes of BRICS-T countries. In contrast, the impact of GDP on environmental performance is found to be negative. In all models, economic growth leads to a decline in environmental performance, thereby revealing the environmental costs of economic activities. Rapid economic growth increases resource consumption and puts pressure on natural areas. These results suggest that the growth process in BRICS-T countries should be carefully managed against environmental impacts. The use of renewable energy is found to have positive impacts on environmental performance and environmental health in particular. This finding suggests that renewable energy investments contribute to environmental sustainability. It also suggests that renewable energy sources can play a critical role for environmental sustainability in BRICS-T countries. In particular, the positive effects of renewable energy use on environmental health and ecosystem vitality suggest that further investments in this area should be encouraged. On the contrary, energy intensity and population density variables have a negative impact on environmental performance. High energy consumption and rapid population growth put significant pressure on environmental resources. This situation makes it difficult to achieve sustainable development goals in BRICS-T countries. In this regard, energy efficiency and sustainable urbanization are of great importance for improving environmental performance.